\documentclass{sf2a-conf2024}
\usepackage{graphicx}
\usepackage{hyperref}
\usepackage[]{natbib}  
\usepackage{epstopdf}
\usepackage{marvosym}
\usepackage{anyfontsize}
\usepackage{transparent}
\usepackage{tikz}
\usetikzlibrary {fadings,patterns}

\def\BibTeX{{\rm B\kern-.05em{\sc i\kern-.025em b}\kern-.08em
    T\kern-.1667em\lower.7ex\hbox{E}\kern-.125emX}}
\bibpunct{(}{)}{;}{a}{}{,}  

\def\inumber{i}                                 

\newcommand{\dotprod}[1]{\cdot}

\def\XX{X}

\def\ipla{{\empty}}                              
\def\istar{\star}                                       

\def\zz{z}
\def\XX{X}

\def\smaxis{a}                                  
\def\norb{n_\istar}                             
\def\spinrate{\Omega}                   

\def\Mbody{M}                                   
\def\Rbody{R}                                   
\def\Mpla{\Mbody_\ipla}                 
\def\Mstar{\Mbody_\istar}                       

\def\pressure{p}

\def\Cp{C_\pressure}

\def\Rpla{\Rbody_\ipla}                 

\def\freq{\sigma}                               
\def\ggravi{g}                                  
\def\frad{\freq_{0}}

\def\temperature{T}

\def\Tsurf{\temperature_\isurf}

\def\height{H}

\def\Rspec{R_{\rm s}}
\def\Rgp{R_{\rm PG}}
\def\Matm{m_{\rm atm}}
\def\rcp{\kappa}

\def\heightsurf{\height_{\isurf}}

\def\ftide{\freq}                                       

\def\isurf{{\rm s}}                             

\def\flux{F}
\def\Fstar{\flux_{\istar}}

\newcommand{\vartide}[1]{\delta #1}

\def\presstide{\vartide{\pressure}}

\def\freso{\ftide_{\rm L}}
\def\disspar{\varepsilon}

\def\torque{\mathcal{T}}                        

\newcommand{\torquei}[1]{\torque_{#1}}

\def\torquez{\torquei{\zz}}

\def\opacity{\alpha_{\rm A}}

\def\dpzero{\delta \pressure_0}
\def\dpquad{\presstide_{2,2}}
\def\dFquad{\delta \flux_{\istar ; 2,2}}

\newcommand{\eq}[1]{Eq.~(\ref{#1})}
\newcommand{\eqs}[2]{Eqs.~(\ref{#1}) and~(\ref{#2})}

\newcommand{\fig}[1]{Fig.~\ref{#1}}

\newcommand{\comments}[1]{}

\pdfpageattr{/Group <</S /Transparency /I true /CS /DeviceRGB>>}
\begin{document}

\TitreGlobal{SF2A 2024}


\title{Thermal tides on rocky planets through a novel fully analytical solution}

\runningtitle{Thermal tides on rocky planets}

\author{P. Auclair-Desrotour}\address{IMCCE, Observatoire de Paris, Université PSL, CNRS, Sorbonne Université, 77 Avenue Denfert-Rochereau, Paris, 75014, France}

\author{M. Farhat$^{1}$}

\author{G. Boué$^{1}$}

\author{R. Deitrick}\address{School of Earth and Ocean Sciences, University of Victoria, Victoria, British Columbia, Canada}

\author{J. Laskar$^{1}$}




\setcounter{page}{237}


\maketitle


\begin{abstract}
Thermal tides are atmospheric tides caused by variations in day-night insolation, similar to gravitational tides but with key differences. While both result in delayed mass redistribution, energy dissipation, and angular momentum exchanges between the planet and its host star, thermal tides can drive a planet's dynamics away from the rotational equilibrium states predicted by classical tidal theory. In this work, we present a novel closed-form solution for the thermotidal response of rocky planets. This general solution is derived from first principles, assuming either dry or moist adiabatic temperature profiles for the planet's atmosphere, and can be readily applied to study the long-term evolution of exoplanets in the habitable zones of their host stars. Despite relying on a small number of parameters, the model successfully captures the key features of the thermotidal torque predicted by General Circulation Models (GCMs). It also accurately predicts Earth's current semidiurnal thermotidal response and provides new insights into the evolution of the length of day during the Precambrian era.  
\end{abstract}



\section{Introduction}
It has long been recognised that Solar irradiation induces atmospheric thermal tides, similar to gravitational forces. These tides arise from the difference in thermal forcing between the day and night sides of the planet, with the dayside heated by the star and the nightside cooling radiatively \citep[e.g.][]{chapman1970atmospheric}. This leads to a denser atmosphere on the nightside, causing global redistribution of mass, akin to oceanic and solid tides. As early as the late 19th century, Lord Kelvin quantified the thermotidal torque, arising from this mass redistribution, on Earth's spin axis using barometric measurements \citep{thomson1882}, finding it to be approximately $7\%$ of the torque from gravitational tides in the Earth's oceans and solid interior \citep[e.g.][]{lambeck1977}, despite the tiny planet's mass fraction constituted by the atmosphere (${\sim} 8.5 \times 10^{-5} \, \%$)\footnote{\url{https://nssdc.gsfc.nasa.gov/planetary/factsheet/earthfact.html}.}. 

Earth's semidiurnal thermal tide is phase-advanced relative to the Sun, with surface pressure oscillations peaking around $09{:}45$ LST (Local Solar Time) and $22{:}00$ LST \citep{dai1999diurnal,schindelegger2014surface,auclair2017atmospheric}, unlike gravitational tides, which are always delayed. The resulting accelerative thermotidal torque increases Earth's spin over time. This effect is also seen on Venus, where it counteracts the torque from solid tides, leading to asynchronous rotation \citep[][]{gold1969atmospheric,ingersoll1978venus,correia2001four,correia2003long,correia2003long2,leconte2015asynchronous}. Thermal tides can similarly induce differential rotation on hot Jupiters \citep[e.g.][]{arras2010thermal,auclair2018semidiurnal,gu2019modeling,lee2020tidal} and determine the rotational states of near synchronous rocky exoplanets \citep[e.g.][]{laskar2004rotation,correia2010tidal,cunha2015spin,leconte2015asynchronous,auclair2017atmospheric,auclair2019generic}. 

\begin{figure}[t]
 \centering
 \includegraphics[width=0.45\textwidth,clip]{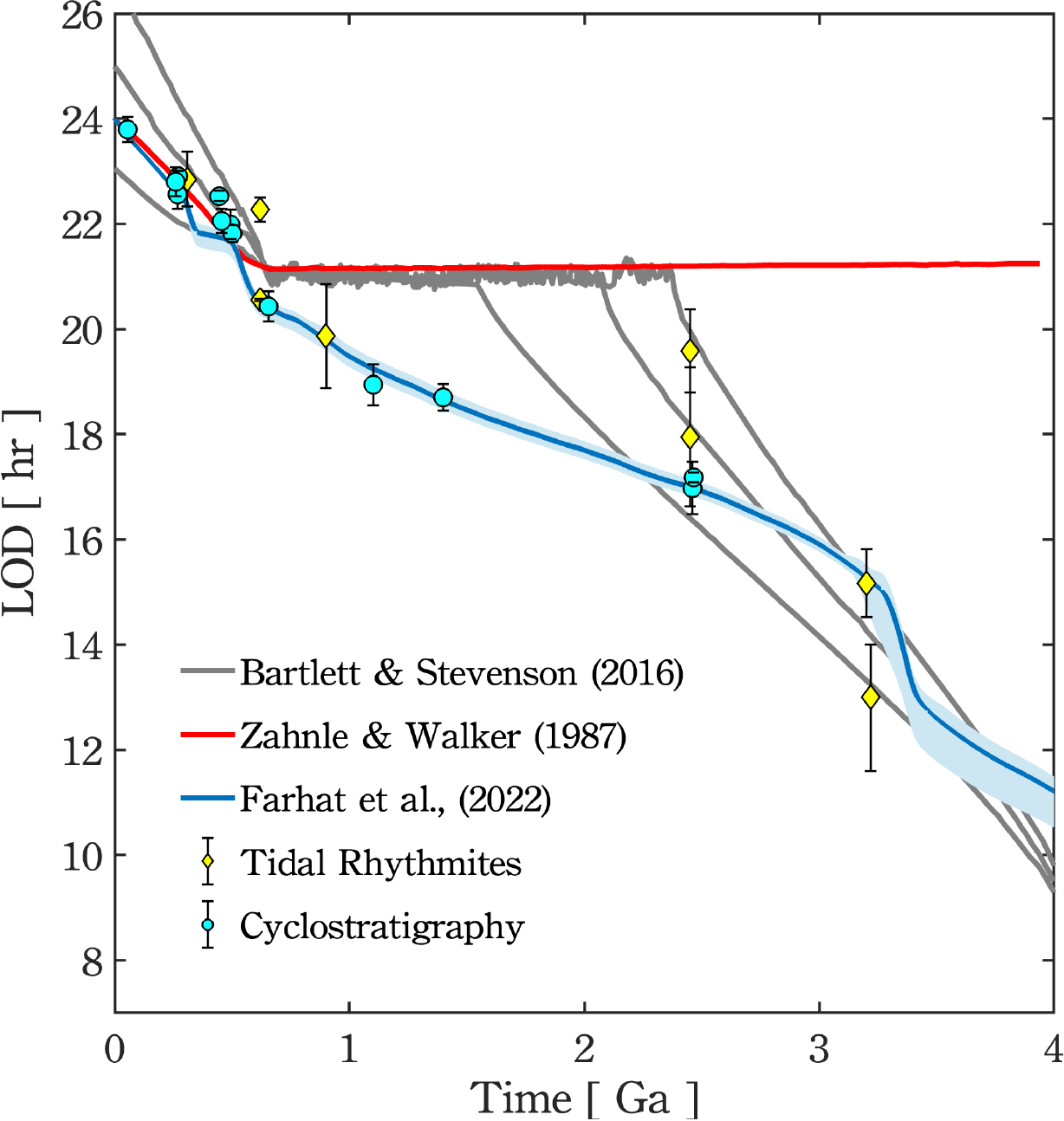}  \hfill
 \includegraphics[width=0.48\textwidth,clip]{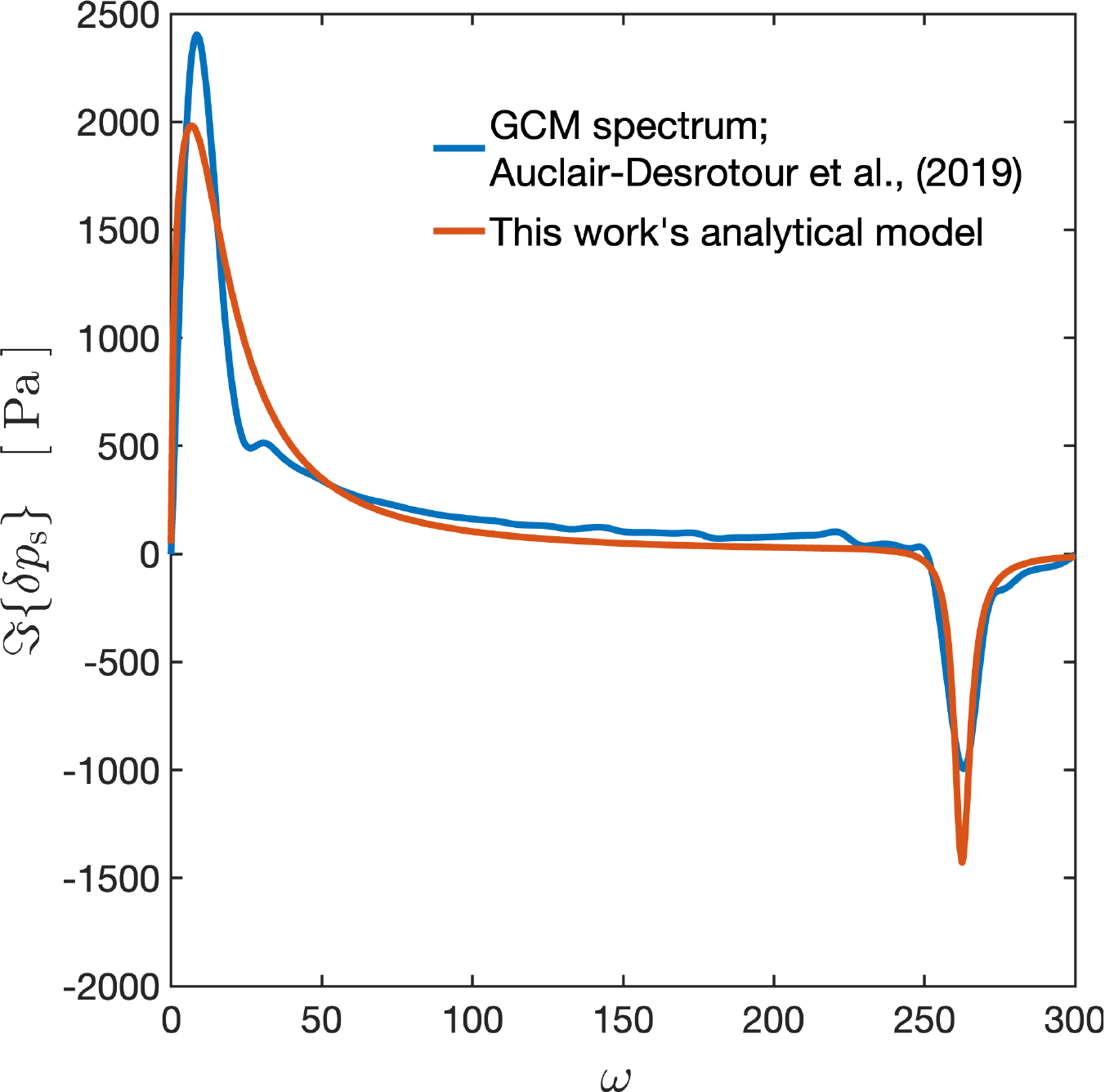}
  \caption{{\bf Left:} Modelled evolution of the Earth's LOD over geological timescales, shown for three different models: (i) \cite{farhat2022resonant}, where the evolution is driven solely by oceanic and solid tidal dissipation; (ii) \cite{zahnle1987constant}, where the resonance of the Lamb wave occurs at a LOD of $ {\sim} 21 \ {\rm hr}$, imposing a rotational equilibrium on Earth; and (iii) \cite{bartlett2016analysis}, who propose equilibrium scenarios with the potential to escape tidal locking. {\bf Right:} Imaginary part of the tidal surface pressure oscillations as a function of the normalised semidiurnal frequency $\omega = \ftide / \left( 2 \norb \right)$ obtained (i) from GCM simulations by \cite{auclair2019generic} (blue line), and (ii) by fitting the analytical model of F24 (\eq{Imdeltap22}) to this numerical result (red line).} 
  \label{auclair-desrotour:fig1}
\end{figure}

While the thermotidal torque on present-day Earth contributes only minimally to the total torque, the atmosphere may have significantly influenced Earth's past rotational history. Like ocean tides, thermal forcing can excite planetary-scale compressibility waves known as Lamb waves \citep[][]{lamb1911atmospheric,bretherton1969,lindzen1972lamb}, analogous to the long-wavelength surface gravity modes in ocean tides. Resonance occurs when the eigenperiod of a Lamb wave matches the semidiurnal period. Given that Earth's current 12-hour semidiurnal period is close to resonance, \cite{zahnle1987constant} proposed that such a resonance occurred during the Precambrian era ($4587-538$~Myr ago), when the length of day (LOD) was around 21 hours. With gravitational tidal torque then about a quarter of today's, they suggested that Earth's LOD may have reached equilibrium and remained stable for hundreds of millions of years, as shown by \fig{auclair-desrotour:fig1} (left panel). Decades after this rotational equilibrium scenario was suggested for the Earth, \cite{bartlett2016analysis} revisited the problem and tested the stability of the equilibrium to temperature changes on Earth. More recently, \cite{mitchell2023mid} and \cite{wu2023day} made use of the growing abundance of geological proxies on the Earth's rotational history to support the scenario originally proposed by \cite{zahnle1987constant}.


In two recent papers, \citep[][hereafter F24 and L24]{farhat2024,laskar2024}, we explore the equilibrium hypothesis using an analytical approach. These studies build on \cite{farhat2022resonant}, who reconstructed the Earth-Moon system's history based solely on oceanic and solid tidal dissipation. In F24, we present a closed-form solution for the tidal response of neutrally stratified atmospheres, dependent on tidal frequency and a few key physical parameters. This model shows strong agreement with General Circulation Models (GCM) simulations from previous work. Applied to Earth, our findings suggest that the resonant amplification of the thermotidal torque was insufficient to counter the torque arising from gravitational tides in the oceans, challenging the tidal locking hypothesis. The second paper, L24, complements F24 by reviewing recent research on the equilibrium problem and geological records. The following discussion focuses on key results from F24, with all figures taken from that work.

\section{New general prescription for thermotidal atmospheric torque}

We consider a star with mass $\Mstar$ hosting a rocky planet of mass $\Mpla$, radius $\Rpla$, surface gravity $\ggravi$, and spin rate $\spinrate$. The planet's atmosphere, assumed to be thin, has a typical thickness given by the pressure scale height at the surface, $\heightsurf = \Rspec \Tsurf / \ggravi$, where $\Rspec$ is the specific gas constant ($\Rspec = \Rgp / \Matm$, with $\Rgp$ being the universal gas constant and $\Matm$ the mean molecular weight), and $\Tsurf$ denotes the globally averaged surface temperature. For a circular orbit of radius $\smaxis$ and mean motion $\norb$, within the planet's equatorial plane, the thermotidal torque reduces to the semidiurnal component,
\begin{equation}
\label{torquez}
\torquez = \sqrt{\frac{6 \pi}{5}} \frac{\Mstar}{\Mpla} \frac{\Rpla^6}{\smaxis^3} \Im \left\{ \dpquad \left( \ftide \right) \right\}.
\end{equation}
Here, $\ftide = 2 \left( \spinrate - \norb \right)$ represents the semidiurnal tidal frequency, $\Im$ refers to the imaginary part of a complex number (with $\Re$ referring to the real part), and $\dpquad$ is the degree-2, order-2 complex component in the spherical harmonic expansion of the tidal pressure oscillation, which captures the frequency-dependent atmospheric response. Additionally, we define the corresponding stellar flux component, $\dFquad$, the resonant frequency of the main forced Lamb wave, $\freso$, and a constant pressure factor, $\dpzero$,
\begin{equation}
\begin{array}{lll}
\displaystyle \dFquad= \frac{\sqrt{30 \pi}}{16} \Fstar, & \displaystyle \freso = \frac{\sqrt{\Lambda_2 \ggravi \heightsurf}}{\Rpla}, &\displaystyle \dpzero = \frac{\rcp \opacity   \dFquad }{\heightsurf \freso},
\end{array}
\end{equation}
where $\Fstar$ is the incident flux at the sub-stellar point, $\Lambda_2$ is the eigenvalue of the Hough function\footnote{Hough functions are the functions used on the sphere to describe the spatial dependence of the tidal response \citep[e.g.][]{wang2016computation}.} that primarily overlaps with the thermal forcing distribution, $\rcp=\Rspec/\Cp$ (with $\Cp$ the heat capacity per unit mass), and $\opacity $ is the fraction of absorbed flux ($0 \leq \opacity \leq 1 $). For simplicity, we assume that $\Lambda_2$ is a constant with value $\Lambda_2 = 11.129$, which corresponds to Earth-like scenarios where $\norb \ll \spinrate$. The variations of $\Lambda_2$ near synchronisation and their potential implications on the atmospheric tidal response are thus neglected. 

Finally, the atmosphere is assumed to be neutrally stratified, meaning that no Archimedean forces are present, and tidal waves are restored solely by horizontal compressibility. Although idealised, this atmospheric structure more realistically describes the convective troposphere of rocky planets than the classical isothermal assumption. Energy dissipation occurs through Newtonian cooling (i.e., linearised radiative cooling), characterised by the frequency $\frad$. Under these conditions, Laplace's tidal equations can be solved analytically within linear theory \citep[e.g.][]{lindzen1967tidal}, yielding 
\begin{equation}
\label{complex_solution_x}
\dpquad = \dpzero \frac{G \left( \XX \right)}{\disspar \left[ \left( \rcp + 1 \right) \XX^2  -1 \right] + \inumber \XX \left( \XX^2 - 1 \right) },
\end{equation}
where $\inumber$ denotes the imaginary unit, $\XX = \ftide / \freso$ is the normalised tidal frequency ($\XX=1$ at resonance), $\disspar = \frad/\freso$ is the damping coefficient, and the transfer function $G \left( \XX \right)$ accounts for soil diffusion effects in case of indirect radiative forcing in the infrared (with $G \left( \XX \right) = 1$ for direct absorption). The imaginary part, $\Im \left\{ \dpquad \right\} $, used in \eq{torquez}, is expressed as 
\begin{equation}
\Im \left\{ \dpquad  \right\} = \dpzero \frac{\XX \left( 1 - \XX^2 \right) \Re \left\{ G \left( \XX \right) \right\} + \disspar \left[ \left( \rcp + 1 \right) \XX^2 - 1 \right] \Im \left\{ G \left( \XX \right) \right\}}{\XX^2 \left( 1 - \XX^2 \right)^2 + \disspar^2 \left[ \left( 1 + \rcp \right) \XX^2 - 1  \right]^2},
\label{Imdeltap22}
\end{equation}
and is plotted in \fig{auclair-desrotour:fig1} after fitting to \cite{auclair2019generic}'s numerical results. For $G \left( \XX \right) =1$ (direct absorption) and $\XX \ll1$ (near-synchronised planets), \eq{complex_solution_x} simplifies to 
\begin{equation}
\dpquad \approx - \frac{\dpzero}{\disspar + \inumber \XX},
\end{equation}
matching earlier analytical or GCM-based predictions for Venus-like planets \citep[][]{ingersoll1978venus,leconte2015asynchronous,auclair2019generic,salazar2024}.

\begin{figure}[t]
 \centering
 \includegraphics[width=0.50\textwidth,clip]{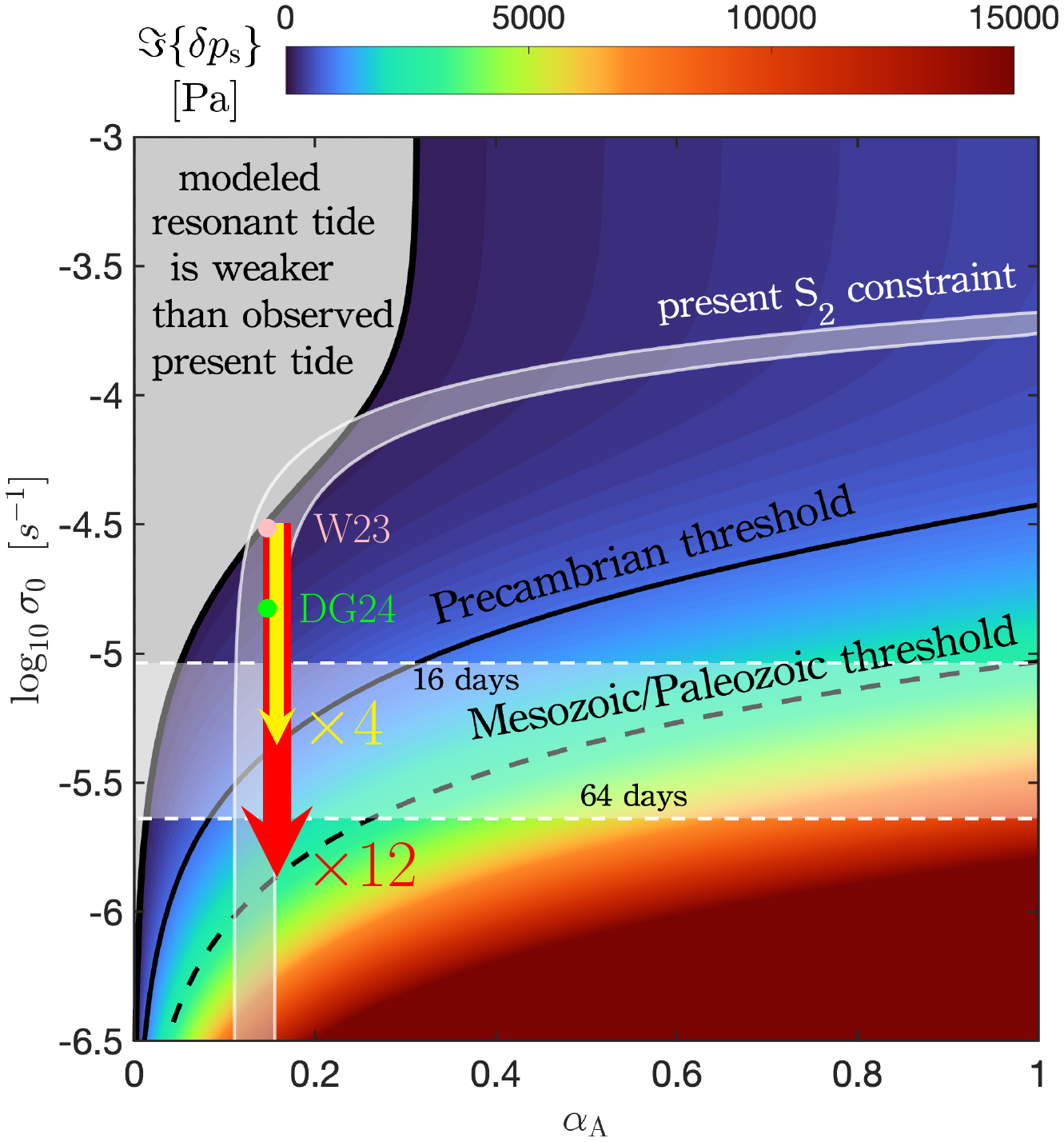}  
  \caption{Maximum reached by the imaginary part of the semidiurnal surface pressure component given by \eq{Imdeltap22} as a function of the atmospheric opacity and radiative cooling frequency. The black solid and dashed lines define, from below, regions where the thermotidal response is sufficient to cancel the gravitational counterpart in the Precambrian ($4587-538$~Myr) and the late Paleozoic/early Mesozoic ($350-250$~Myr). The shaded areas indicate the contraints on $\opacity$ and $\frad$ derived from observations and GCM simulations, along with the corresponding radiative timescales. The yellow and red arrows indicate the amplification factors needed to reach the Precambrian and Mesozoic/Paleozoic thresholds, respectively. The pink and green dots represent the GCM simulations of \cite{wu2023day}, Fig.~S4 (W23), and of \cite{deitrick2024}, Fig.~1 (DG24), respectively. Figure adapted from F24, Fig.~5.  } 
  \label{auclair-desrotour:fig2}
\end{figure}

\section{Model predictions for Earth's past evolution}

\paragraph{Observationally consistent estimate of present-day torque} The analytical solution derived from \eqs{torquez}{Imdeltap22} allows for a broad exploration of the parameter space. When applied to Earth, it offers fresh insights into the equilibrium hypothesis. Notably, its prediction of the present-day thermal torque aligns well with estimates obtained from global barometric measurements \citep[e.g.][]{schindelegger2014surface}. 

\paragraph{Late resonance crossing} Figure~\ref{auclair-desrotour:fig2} illustrates the maximum value of the imaginary part of the semidiurnal surface pressure component, plotted against atmospheric opacity and radiative cooling frequency for direct forcing ($G \left( \XX \right) =1$ in \eq{Imdeltap22}). Observational and numerical constraints suggest that the minimum thermotidal torque required to balance gravitational torque during the Precambrian could only have been achieved with cooling timescales near the upper limits predicted by GCM simulations. The model in F24, however, suggests that resonance more likely occurred during the Mesozoic or the Paleozoic eras, when oceanic tidal dissipation was stronger \citep[][]{zahnle1987constant,farhat2022resonant}. The thermotidal response at resonance appears insufficient to reach the Mesozoic/Paleozoic threshold shown in \fig{auclair-desrotour:fig2}. 

\paragraph{Limited resonant amplification} It is worth noting that the solution provided by \eq{Imdeltap22} likely overestimates the strength of the thermal tide at resonance, as non-linear dissipative processes are expected to damp tidal waves more effectively than the Newtonian cooling model used here. GCM simulations tend to confirm this, predicting resonant amplification factors of around $1.5$ \citep[][]{wu2023day,deitrick2024}, whereas factors of approximately $4$ or $12$ would be required to reach the Precambrian or Mesozoic/Paleozoic thresholds, respectively. 

\paragraph{Symmetry-breaking effect of soil diffusion} The resonance of the Lamb wave is characterised by two peaks: one accelerative, potentially responsible for tidal locking, and one braking, which slows the planet's spin similarly to gravitational tides. In GCM simulations, the accelerative peak is significantly weaker than the braking one \citep[][]{auclair2019generic,wu2023day,deitrick2024}, further arguing against the equilibrium hypothesis. This asymmetry can be captured in the analytical model by incorporating the effects of soil diffusion and indirect thermal forcing in the infrared by the ground, as shown in \fig{auclair-desrotour:fig1} (right panel). 

\section{Conclusions}

By applying classical tidal theory to an atmosphere with neutral stratification and radiative cooling, we derived a new solution describing the frequency-dependent thermotidal torque exerted on a rocky planet's spin axis. This solution reflects key features previously identified in GCM simulations and has been validated against observational data. When combined with GCM results, the model's main predictions suggest that tidal locking likely never occurred on Earth. The derived expression for the thermotidal torque is particularly valuable for exoplanet studies investigating the tidal spin evolution of rocky planets with atmospheres. 

\begin{acknowledgements}
This work has been supported by the French Agence Nationale de la Recherche (AstroMeso ANR-19-CE31-0002-01) and by the European Research Council (ERC) under the European Union's Horizon 2020 research and innovation program (Advanced Grand AstroGeo-885250). This work was granted access to the HPC resources of MesoPSL financed by the Region Île-de-France and the project Equip@Meso (reference ANR-10-EQPX-29-01) of the programme Investissements d'Avenir supervised by the Agence Nationale pour la Recherche. 
\end{acknowledgements}

\bibliographystyle{aa}  
\bibliography{Auclair-Desrotour_S13} 

\end{document}